\documentclass[aps,prd,preprint,floatfix,nofootinbib,longbibliography,a4paper]{revtex4-1}
\pdfoutput=1
\usepackage{color}
\usepackage{graphicx}
\usepackage{textcomp}
\usepackage{subfigure}
\usepackage{amsmath,amssymb,array}
\usepackage{enumerate}
\usepackage{hhline}
\usepackage[normalem]{ulem}
\usepackage{soul}
\usepackage{cancel}
\newcommand{\mathsym}[1]{{}} 
\def\lsim{\:\raisebox{-1.1ex}{$\stackrel{\textstyle<}{\sim}$}\:}
\def\gsim{\:\raisebox{-1.1ex}{$\stackrel{\textstyle>}{\sim}$}\:}
\def\vev#1{\left\langle #1\right\rangle}

\newcommand{\beqa}{\begin{eqnarray}}
\newcommand{\eeqa}{\end{eqnarray}}
\newcommand{\be}{\begin{equation}}
\newcommand{\ee}{\end{equation}}
\newcommand{\ba}{\begin{array}} 
\newcommand{\ea}{\end{array}}

\begin{document} 
\vspace*{0.5cm}
\title{Inflation and long-range force from clockwork $D$-term}
\bigskip
\author{Anjan S. Joshipura}
\email{anjan@prl.res.in}
\author{Subhendra Mohanty}
\email{mohanty@prl.res.in}
\author{Ketan M. Patel}
\email{kmpatel@prl.res.in}
\affiliation{Physical Research Laboratory, Navarangpura, Ahmedabad-380 009, India \vspace*{1cm}}

\begin{abstract}
Cosmic inflation driven by the vacuum energy associated with the $D$-term of a supersymmetric abelian gauge group and a possible existence of long-range force mediated by an ultra-light gauge boson $Z^\prime$ are two extreme examples of models based on extra $U(1)$ symmetries. Large vacuum energy sets the scale of inflation while the scales of long-range forces induced by anomaly free extra gauged $U(1)$ symmetries are constrained by neutrino oscillations, binary pulsar timings and invisible neutrino decay. There exists a difference of about 40 orders of magnitude between the scales of these two. Also, gauge couplings associated with the long-range forces are very small compared to the standard model couplings and the one required for inflation. We propose a framework based on clockwork mechanism in which these vastly different scales and associated new physics can coexist without invoking any arbitrarily small or large parameter in the fundamental theory. A chain of $U(1)$ is introduced with characteristic nearest-neighbour interactions. A large $D$-term introduced at one end governs the dynamics of inflation. $Z^\prime$ is localized on the other end of the chain, and it can be massless or can get naturally suppressed mass. The standard model fields can be charged under one of the intermediate $U(1)$ in the chain to give rise to their small effective coupling $g^\prime$ with $Z^\prime$. Constraints on $g^\prime$ and $M_{Z^\prime}$ are discussed in the context of the long-range forces of type $L_\mu - L_\tau$, $L_e - L_\mu$ and $B-L$. These, along with the inflation observables, are used to constraint the parameters of the underlying clockwork model. 
\end{abstract}

\maketitle

\section{Introduction}
There exist a large number of well-motivated gauged extensions of the Standard Model (SM) containing an extra $U(1)$ group. These are proposed (a) on phenomenological grounds like explaining anomaly found in the muon anomalous magnetic moment \cite{Baek:2001kca} (see also \cite{Lindner:2016bgg} for a review) or as explanation of the universality violation observed in the $B$ meson decays \cite{Altmannshofer:2016jzy}, (b) on cosmological grounds such as need to explain the dark matter \cite{Holdom:1985ag,Hooper:2012cw,Fabbrichesi:2020wbt}, to provide secret interactions between sterile neutrinos of eV masses \cite{Hannestad:2013ana,Dasgupta:2013zpn} to suppress their cosmological production in the early universe etc., (c) as a theoretical framework for the successful description of the inflation in the context of supersymmetric versions of the SM \cite{Stewart:1994ts,Binetruy:1996xj,Halyo:1996pp}, and (d) to provide a simple description of the long-range ``fifth force" \cite{Fayet:1986vz,Fayet:1989mq} if it exists. Examples of such $U(1)$ are difference of any two of the leptonic charges $L_{e,\mu,\tau}$  \cite{He:1990pn,He:1991qd,Foot:1994vd} or an unbroken or mildly broken $B-L$ symmetry \cite{Heeck:2014zfa}.

Many extensions in category (a) and (b) need a very light gauge boson typically in the mass range eV-MeV. The models of the $D$-term inflation \cite{Stewart:1994ts,Binetruy:1996xj,Halyo:1996pp} use the Fayet-Illiopoulos (FI) term \cite{Fayet:1974jb} which can be written when the gauge symmetry is $U(1)$. A large value for the FI parameter $\xi  \sim 10^{32}$ GeV$^2$ leads to inflation in the early universe driven by an almost flat potential. Extensions in the category (d) correspond to an entirely different parameter range. If the first generation fermions are charged under the extra $U(1)$ then the induced long-range forces are constrained by the fifth force experiments \cite{Touboul:2017grn} or by the precision tests of gravity \cite{Kapner:2006si,Schlamminger:2007ht}. These experiments constrain the couplings of electrons to the light gauge boson $M_{Z^\prime}$ and are not sensitive to the neutrino couplings. Constraints on the masses and couplings for the range of length $\gsim 0.1\, {\rm eV}^{-1}$ follow from these experiments and restrict the coupling $g^\prime$ to be $\lsim 10^{-25}$. If $U(1)$ group distinguishes between leptonic flavours then the long-range forces generated by electrons from the earth, Sun, Galaxies etc. induce the matter effects in neutrino oscillations \cite{Joshipura:2003jh,Grifols:2003gy}. This effect can suppress the observed neutrino oscillations for a range in the gauge boson mass $M_{Z^\prime}$ and coupling $g^\prime$. Terrestrial experiments, as well as astrophysical and cosmological considerations, constrain the allowed  $M_{Z^\prime}$-$g^\prime$ parameter space.  It is found that there exists a region of parameters for which the $Z^\prime$ induced potential can be comparable to the Wolfenstein potential induced by the charged current interaction in the SM. This happens \cite{Smirnov:2019cae}  for approximate ranges $M_{Z^\prime}\sim 10^{-17}$-$10^{-14}$ eV and $g^\prime \sim 10^{-27}$-$10^{-25}$. This implies a strong hierarchy $M_{Z^\prime} / \sqrt{\xi} \sim 10^{-40}$ between the allowed ultra-light mass and the inflation scale. Considering that the scales and parameters associated with the SM are much larger than $M_{Z^\prime}$ and $g^\prime$, it is natural to seek a theoretical explanation of their smallness.

It was pointed out by Fayet \cite{Fayet:1984jt,Fayet:1990wx} (see also \cite{Fayet:2016nyc,Fayet:2017pdp,Fayet:2018cjy}) that the presence of a FI term allowed in case of the supersymmetric gauge theories can be used to relate the inflation scale $\sqrt{\xi}$ to a  very small gauge coupling $g^\prime$. Consider a simple supersymmetric gauge theory based on a $U(1)$ group containing two oppositely charged superfields $\phi_{\pm}$. The scalar potential of this theory includes the following $D$-term contribution.
\be \label{vd}
V_D=\frac{1}{2} (\xi-g^\prime(|\phi_+|^2 - |\phi_-|^2))^2~.\ee
This simple potential is used to drive inflation when it is supplemented with a gauge singlet superfield $X$ - the inflaton and an $F$-term coming from a superpotential $\lambda X \phi_+\phi_-$. A large value of the inflaton field in the early universe leads to a supersymmetry breaking and $U(1)$ preserving minimum of $V_D$ with a value $\frac{1}{2} \xi^2$ at the minimum. For vanishing $X$ field which occurs after inflation, the $V_D$ has a supersymmetry preserving but the gauge symmetry breaking minimum with $g^\prime |\vev{\phi_+}|^2=\xi$.  The above $D$-term leads to a scalar mass term $\mu^2=g^\prime \xi$\footnote{More precisely this will be a $D$-term contribution to the mass of the real part of $\phi_+$ when $g^\prime \vev{\phi_+}^2=\xi$.}. Requiring that  this mass parameter is less  than the typical supersymmetry breaking scale $\sim$ TeV gives  a small $g^\prime \leq {\rm TeV}^2/\xi \sim 10^{-26}$ \cite{Fayet:2017pdp,Fayet:2018cjy}. Thus a large inflation scale relates to very small value of $g^\prime$. While small value of $g^\prime$ follows in this simple $U(1)$ example, it still cannot describe the long range forces. The $U(1)$ gauge boson in this case  acquires a mass $M_{Z^\prime}^2= g^{\prime 2} |\vev{\phi_+}|^2=g^\prime \xi=\mu^2\sim {\rm TeV}^2$ which leads to a very short range potential. This is an artefact of the use of the SM singlet fields $\phi_\pm$ for breaking the $U(1)$ symmetry. As shown by Fayet \cite{Fayet:1984jt}, it is possible to obtain ultra-light  or even a massless $M_{Z'}$ \cite{Fayet:1990wx}, a small $g^\prime$ and the flat potential required for the inflation to  start by using the  SM non-singlet fields to break the  $U(1)$ symmetry\footnote{A specific example based on the $SU(5)\times U(1)$ group proposed in \cite{Fayet:1984jt}, leads to a mass relation $M_{Z'}=\frac{g^\prime}{g} M_W$  leading to ultra-light gauge boson for very small $g'$.}.

As an alternative to the above setup, we propose a series of $N+1$ gauged $U(1)_i$ groups ($i=0,1,...,N$) based on the clockwork (CW) mechanism \cite{Kaplan:2015fuy,Giudice:2016yja}. $N$ chiral superfields are introduced, each of which couples to only two adjacent $U(1)$ in the chain leading to characteristic nearest-neighbour interactions. FI term is introduced only for the $U(1)$ at the $N^{\rm th}$  site. This leads to inflation in a manner described above. The corresponding gauge coupling is of ${\cal O}(1)$. All the $U(1)$ symmetries, except a linear combination of them, get broken at the minimum, but the breaking scales are hierarchically related to the FI parameter $\xi$. Specifically, the $U(1)_i$ is broken at a scale $\sim q^{\frac{N-i}{2}}\sqrt{\xi}$, where $q$ being the $U(1)$ charge carried by chiral superfields which induce the symmetry breaking. The remaining $U(1)$ symmetry is broken by introducing another pair of chiral superfields which couple to one of the intermediate $U(1)$ in the chain. The localization of chiral superfields away from $U(1)_0$ leads to an explanation of a large hierarchy between the scales of inflation and the mass of the gauge boson mediating long-range force. The SM fields also interact with one of the intermediate $U(1)$ with ${\cal O}(1)$ gauge coupling. Exponentially small coupling with lightest gauge boson is then obtained in a manner used to describe the mini-charged particles within the standard CW frameworks \cite{Giudice:2016yja,Lee:2017fin}.

We introduce the basic framework of CW $D$-term in the next section. Inflation driven by the $D$-term along with the implications on inflationary observables is discussed in section \ref{sec:inflation}. In section \ref{sec:constraints}, we collect various laboratory, astrophysical and cosmological constraints on the popular class of long-range forces and discuss their consequences on the parameters of the CW framework. We summarize the study in section \ref{sec:concl}.

\section{Framework}
\label{sec:framework}
\begin{figure}[t!]
\centering
\subfigure{\includegraphics[width=0.98\textwidth]{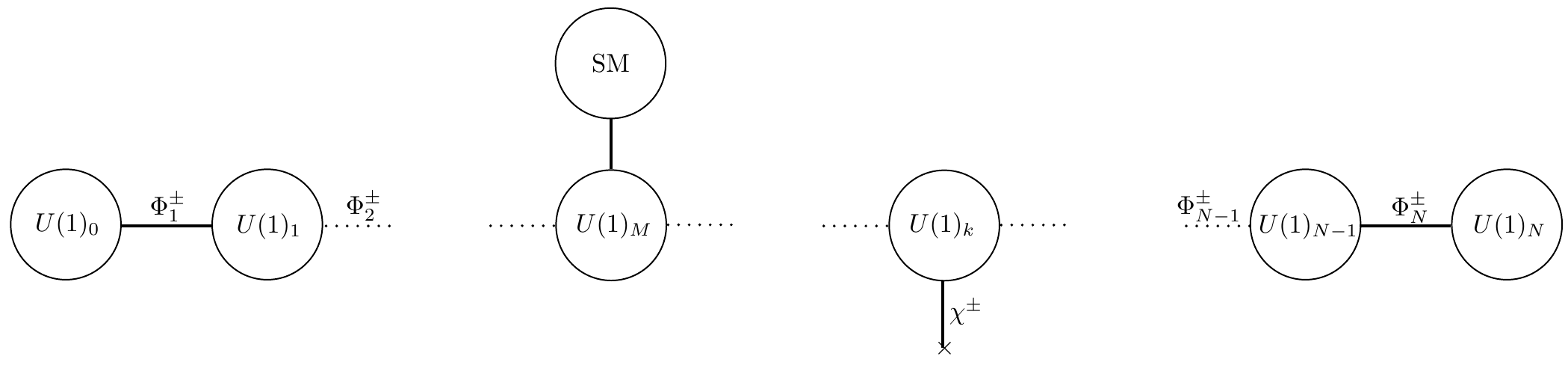}}
\caption{Schematic presentation of the clockwork model used in the present study.}
\label{fig1}
\end{figure}
The framework of multiple $U(1)$ we discuss here is based on the clockwork constructions discussed in \cite{Giudice:2016yja,Lee:2017fin,Ahmed:2016viu}. Consider a chain of $N+1$ supersymmetric $U(1)_i$, with $i=0,1,...,N$. A vector superfield  ${\cal V}_i$ of $U(1)_i$ contains a vector filed $\hat{V}_{i,\mu}$, a pair of weyl fermions $\lambda_i$, $\lambda_i^\dagger$ and auxiliary field $D_i$. The supersymmetric Lagrangian involving gauge fields is given by
\beqa \label{L_gauge}
{\cal L}_{\rm gauge} & = & \sum_{i=0}^{N}\, \left(-\frac{1}{4} F_i^{\mu \nu} F_{i,\mu \nu} + i \lambda^\dagger_i \overline{\sigma}^\mu \partial_\mu \lambda_i + \frac{1}{2} D_i^2 \right)\,,
\eeqa
where $F_{i,\mu \nu} = \partial_\mu \hat{V}_{i,\nu} - \partial_\nu \hat{V}_{i,\mu}$. One can further include a gauge and supersymmetry invariant Fayet-Iliopoulos (FI) term for each $U(1)$. Here, we assume that only $U(1)$ at the $N^{\rm th}$ site possesses such a term.
\be \label{FI}
{\cal L}_{\rm FI} = - \xi\, D_N\,. 
\ee
The assumption of having only one vanishing FI term is technically natural \cite{Fischler:1981zk} as the trace of each $U(1)_i$ factor is individually zero.
We then consider $N$ pairs of chiral superfields $ \Phi^\pm_i$ (with $i=1,...,N$) charged under $U(1)_{i-1} \times U(1)_i$ with charges $( \mp q_{i-1}, \pm 1)$. These fields are chargeless under all the other $U(1)$ in the chain.   The schematic presentation of the model is displayed in Fig. \ref{fig1}. We also consider a chiral superfield $X$ neutral under all the $U(1)$ groups. The relevant superpotential considered in the underlying framework is
\be \label{W}
W = \sum_{i=1}^N\, \lambda_i\, X\Phi_i^+ \Phi_i^- \,.
\ee
The other terms in the potential may be forbidden by imposing additional symmetries\footnote{For example, an $R$-symmetry under which $W \to e^{i \alpha } W$ and $ X \to e^{i \alpha } X$ along with a $Z_2$ symmetry under which only $X$ and $\Phi_i^+$ are odd can forbid all the other terms in $W$.}. The gauge interaction between chiral and vector superfields contains the following interaction term between the scalars $\phi_i^\pm$ residing in $\Phi_i^\pm$ and $D_i$ in ${\cal V}_i$.
\be \label{L_int}
{\cal L}_{\rm int} \supset \sum_{i=1}^N\, \left(g_i D_i - g_{i-1} q_{i-1} D_{i-1}\right)\,G_i \,, \ee
where $G_i = |\phi_i^+ |^2 - |\phi_i^- |^2$ and $g_i$ is the gauge coupling corresponding to $U(1)_i$. 

Elimination of the auxiliary fields from Eqs. (\ref{L_gauge},\ref{FI},\ref{L_int}) using the equations of motion implies
\be \label{D_eq}
D_j = \delta_{jN} \xi - g_j\, (G_j - q_j\, G_{j+1})\,, \ee
where $j=0,1,...,N$ and $G_0=G_{N+1}= 0$. Substituting this solution in Eqs. (\ref{L_gauge},\ref{FI},\ref{L_int}) leads to the following $D$-term scalar potential
\be \label{VD}
V_D = \frac{1}{2} \xi^2 - \xi g_N G_N + \sum_{i=1}^N \left(\frac{1}{2} (g_i^2+g_{i-1}^2 q_{i-1}^2)\, G_i^2 - g_{i-1}^2 q_{i-1} G_i G_{i-1} \right)\,.
\ee
This together with the $F$-term potential derived from Eq. (\ref{W}),
\be \label{VF}
V_F = \sum_{i=1}^N\, \lambda_i^2 |X|^2 \left(|\phi_i^+|^2 + |\phi_i^-|^2\right) +  \sum_{i,j=1}^N\, \lambda_i \lambda_j\ (\phi_i^+\phi_i^-)^* (\phi_j^+\phi_j^-)\,, \ee
gives the complete scalar potential of the underlying framework, $V = V_F + V_D$.

\subsection{Symmetry breaking}
It is seen from Eqs. (\ref{VD},\ref{VF}) that  the potential has a minimum at $|\phi^\pm_i|=0$  when $|X|^2 \ge g_N \xi / \lambda_N^2$. Consequently, the gauge symmetry is unbroken but supersymmetry gets broken by $D_N = \xi$. This implies an almost flat potential  $V \simeq \frac{1}{2} \xi^2$ with non-zero slope provided by loop corrections as it will be described  in section \ref{sec:inflation}.  When $X$ rolls down to its minimum, the vacuum expectation values (VEV) of other fields are determined by the minimization of the potential $V$. The minimum of $V_D$ corresponds to 
\be \label{G_j}
G_j = \frac{1}{ g_j^2 + g_{j-1}^2 q_{j-1}^2}\, \left( \delta_{jN}\, g_N \xi  + g_j^2 q_j\, G_{j+1} + g_{j-1}^2 q_{j-1}\, G_{j-1}\right)\,,\ee
for $j=1,...,N$. These $N$ equations can be iteratively solved to find minimum for $|\phi^\pm_j|^2$ along with the minimization of $V_F$.

For simplicity, we now assume $g_0 = g_1 = ... = g_{N-1} \equiv g_N$ and $q_0 = q_1 = ...= q_{N-1} \equiv q$. The absolute minimum of full potential $V$ then occurs for the following:
\be \label{minimum}
|X| = 0\,, ~ |\phi^-_j|^2  = 0\,,~|\phi^+_j|^2= \frac{{\cal N}_N^2}{{\cal N}_{j-1}^2}\, q^{N-j}\, \frac{\xi}{g_N} \equiv v_j^2\,, \ee
where 
\be \label{N_i}
{\cal N}_i^2 = \frac{1}{1+q^2+q^4+....+q^{2i}} = \frac{1-q^2}{1- q^{2(i+1)}}\,. \ee
At this minimum, $D_j$ for $j=0,1,...,N$ are given by 
\be \label{D_j}
D_j= {{\cal N}^2_N}\, q^{N+j}\, \xi.
\ee 
Consequently, all the $D_j$ are non-zero and the supersymmetry is broken in each $U(1)$ sector in the true minimum. The potential $V$ at the minimum is given by 
\be \label{vmin}
V_{\rm min}=\frac{1}{2}\, {\cal N}^2_N\, q^{2N} \xi^2\,.\ee
By choosing large $N$ and $q < 1$, the supersymmetry breaking effects arising from this minimum can be made small. The vacuum structure given in Eq. (\ref{minimum}) breaks all the $U(1)$ individually but leaves a linear combination $U(1)^\prime$ unbroken. The corresponding generator $T^\prime$ can be identified in terms of $U(1)_i$ generators $T_i$ as
\be \label{tp}
T^\prime = {\cal N}_N\,  \sum_{i=0}^N\, q^{i}\,T_i\,. \ee
It is seen that $U(1)^\prime$ is dominantly localized near $U(1)_0$ when $q<1$.

The masses of $N+1$ gauge bosons can be obtained from the kinetic term of $\phi_i^+$ using the following expression of covariant derivative:
\be \label{}
D^\mu \phi_i^+ = \left(\partial^\mu + i g_i \hat{V}_i^\mu - i g_{i-1} q_{i-1} \hat{V}_{i-1}^\mu \right)\, \phi^+_i\,. \ee
We find the gauge bosons mass term
\be \label{GBM}
{\cal L}_{m} = \left( \hat{M}_{V}^2 \right)_{ij}\, \hat{V}^\mu_i \hat{V}_{j,\mu}\,,
\ee
where $i,j = 0,1,...,N$ and the elements of $(N+1) \times (N+1)$ matrix are given by
\be \label{M2V}
\left(\hat{M}_{V}^2 \right)_{ij}= \begin{cases} g_i^2 \left(v_i^2 (1-\delta_{i0})+ q_i^2 v_{i+1}^2 \right)& \mbox{for } j=i \\ - g_i g_j q_j v_i^2  & \mbox{for } j=i - 1 \\
- g_j g_i q_i v_j^2  & \mbox{for } j=i + 1 \\  0 & \mbox{otherwise}\end{cases}\,
\ee

For $g_0 = g_1 = ... = g_{N-1} = g_N$ and $q_0 = q_1 = ...= q_{N-1} = q$, the gauge boson mass matrix, at the leading order in $q$, is then given by
\be \label{M2V_matrix}
\hat{M}_{V}^2  \simeq {\cal N}_N^2\, g_N\, \xi \left( \ba{ccccccccc} 
q^{N+1} & -q^N & 0 & 0 &~ .~ &~ .~ & ~.~ & 0 & 0 \\
-q^N & q^{N-1} &- q^{N-1} & 0 & . & . & . & 0 & 0 \\
0 & - q^{N-1} &  q^{N-2} & -q^{N-2} & . & . & . & 0 & 0 \\
0 & 0 & -q^{N-2} & q^{N-3} & . & . & . & 0 & 0 \\
. & . & . & . & . & . & . & . & . \\
. & . & . & . & . & . & . & . & . \\
. & . & . & . & . & . & . & . & . \\
0 & 0 & 0 & 0 & . & . & . & q & -q \\
0   & 0 & 0 & 0 & . & . & . & -q & 1  \ea\right)\,.
\ee
This matrix admits a massless state which is a specific linear combination of all the $\hat{V}^\mu_i$ states given by 
\be \label{massless}
V_0^\mu\equiv  {\cal N}_N\,\sum_{i=0}^{N} q^i \hat{V}^\mu_i\,. \ee
The other mass eigenstates  $V^\mu_j$ can be determined by an approximate diagonalization of Eq. (\ref{M2V_matrix}). Defining an orthogonal transformation 
\be \label{Rotation}
\hat{V}_i^\mu = \sum_{j=0}^{N}\, {\cal R}_{ij}\, V^\mu_j\,, 
\ee
such that ${\cal R}^T \hat{M}_{V}^2 {\cal R} \equiv {\rm Diag.}(0,M_{V_1}^2,....M_{V_N}^2)$, we find the following form of $ {\cal R}$ at the leading order.
\be \label{R_matrix}
{\cal R} \approx  \left( \ba{ccccccccc} 
1-\frac{q^2}{2} & -q & 0 & 0 &~ .~ &~ .~ & ~.~ & 0 & 0 \\
q & 1-q^2 & - q & 0 & . & . & . & 0 & 0 \\
q^2 & q &  1-q^2 & -q & . & . & . & 0 & 0 \\
q^3 & q^2 & q & 1-q^2 & . & . & . & 0 & 0 \\
. & . & . & . & . & . & . & . & . \\
. & . & . & . & . & . & . & . & . \\
. & . & . & . & . & . & . & . & . \\
q^{N-1} & q^{N-2} & q^{N-3} & q^{N-4} & . & . & . & 1-q^2 & -q \\
q^N & q^{N-1} & q^{N-2} & q^{N-3} & . & . & . & q & 1-\frac{q^2}{2}  \ea\right)\,.
\ee
The zeros in ${\cal R}$ denote analytically approximated values. Numerically, for exact ${\cal R}$, we find their magnitudes non-zero but more suppressed than the other elements present in the corresponding row. The masses of physical states, $V^\mu_j$ with $j=1,...,N$, are obtained as
\be \label{MV_j}
M_{V_j} \simeq q^{(N-j)/2}\, \sqrt{g_N\,\xi}\,,
\ee
at the leading order.

A linear combination of $\hat{V}_{j,\mu}$, identified as $V_{0,\mu} \equiv Z_\mu^\prime$, is massless as a consequence of unbroken $U(1)^\prime$. A small mass for $Z^\prime$ can be generated by introducing an additional pair of chiral superfields ${\bf \chi}^{\pm}$. Let's assume that such a pair is charged under some $U(1)_k$ in the chain and neutral under the rest of $U(1)$. The VEVs of the scalar components of $\chi^{\pm}$ then give an additional contribution to the gauge boson mass term Eq. (\ref{GBM}) given by
\be \label{delta_M2V}
\delta {\cal L}_m = g_k^2\ v_\chi^2\, \hat{V}^\mu_k \hat{V}_{k,\mu}\,,
\ee
where $ v^2_\chi = \left( \langle \chi^+\rangle^2 + \langle \chi^-\rangle^2\right)$. For $v_\chi^2 < v_k^2$, the leading order diagonalization of the modified gauge boson mass matrix is still given by the orthogonal matrix ${\cal R}$ given in Eq. (\ref{R_matrix}). Using this, one obtains in the physical basis
\be \label{delta_M2V_physical}
\delta {\cal L}_m \simeq g_N^2\ v_\chi^2\, \sum_{j=0}^{k+1} q^{2|k-j|}\, V^\mu_j V_{j,\mu}\,.\ee
As a result, mass of the $j^{\rm th}$ gauge boson gets shifted by $\delta M_{V_j} \simeq g_N v_\chi q^{|k-j|}$ for $j \le k+1$. This shift gives mass to the $Z^\prime$ given by 
\be \label{MZp}
M_{Z^\prime} \simeq g_N\, q^k\, v_\chi\,.
\ee
Evidently, an ultra-light $Z^\prime$ can be obtained by either localizing $\chi$ near $U(1)_0$ and choosing tiny $v_\chi$ or taking $k$ close to $N$ and appropriate $v_\chi$ respecting the constraint $v_\chi \lesssim v_k$.

\subsection{Coupling with the Standard Model fields}
Assume that some of the SM fields are charged under $U(1)_M$ in the CW chain such that $0 < M \le N$ (see Fig. \ref{fig1}). The coupling between the $U(1)_M$ current $J^\mu$ of these fields  and the physical gauge boson of $U(1)_j$ is then given by a neutral current interaction term
\be \label{L_NC}
{\cal L}_{\rm NC} = g\, J^\mu \hat{V}_{M,\mu} \simeq g\,J^\mu\, \sum_{j=0}^{M+1}\, q^{|M-j|}\,  V_{j,\mu}\,,
\ee
where $g\equiv g_M$ and the second equality follows from Eqs. (\ref{Rotation},\ref{R_matrix}). For $j > M+1$, one finds the coefficient much smaller than $q^{|M-j|}$ and therefore we neglect these terms and truncate the sum at $j=M+1$. The effective coupling of $Z^\prime$ boson with the $J_\mu$  is then given by
\be \label{g'}
g^\prime = g\, q^{M}\,.
\ee
Hence, $g^\prime$ can be made exponentially suppressed choosing appropriately large $M$ and $q < 1$. 

\begin{figure}[t!]
\centering
\subfigure{\includegraphics[width=0.50\textwidth]{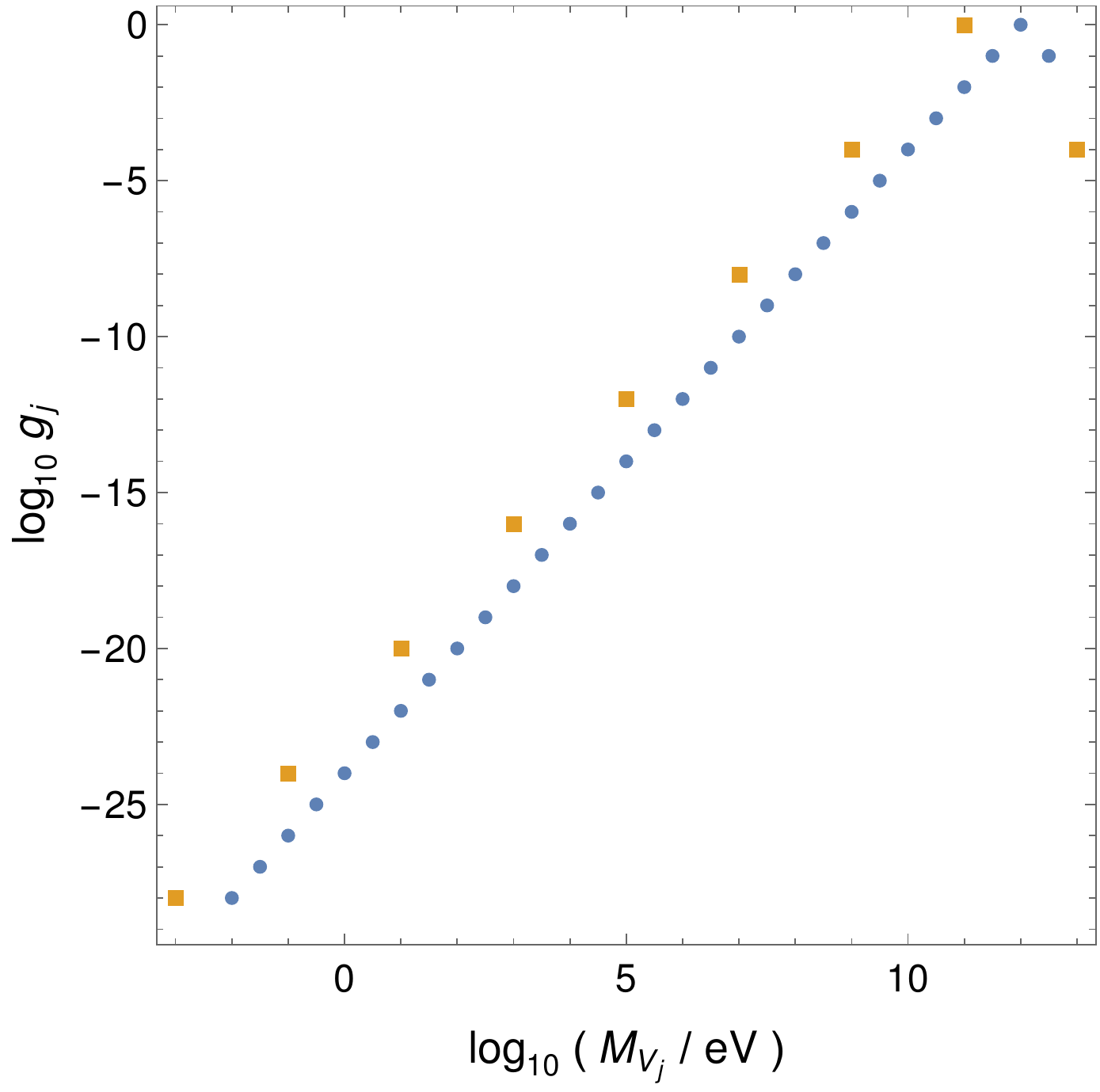}}
\caption{Mass $M_{V_j}$ and coupling $g_j$ of the gauge boson mode $V_j$ ($j=1,2,...,M+1$) as obtained from Eqs. (\ref{MV_j},\ref{L_NC}) for $g=1$ and $\sqrt{g_N \xi} = 10^{16}$ GeV. The blue dots (orange squares) are for $N=55$, $q=0.1$ ($N=15$, $q=10^{-4}$) and the corresponding value of $M$ determined from Eq. (\ref{M}) is 29 (8) for $M_{V_M} = 1$ TeV. }
\label{fig2}
\end{figure}

As seen from Eq. (\ref{L_NC}), the gauge boson with the strongest coupling with to $J_\mu$ is  $V_M$. If $g \sim {\cal O}(1)$, the mass $\sim q^{\frac{N-M}{2}}\sqrt{g_N \xi}$ of this gauge boson is dominantly constrained from the direct search experiments depending on the exact nature of $J^\mu$. Therefore, $M$ can be determined from Eq. (\ref{MV_j}) as
\be \label{M}
M = N - 2 \frac{\ln \left(M_{V_M}/\sqrt{g_N \xi} \right)}{\ln q}\,,\ee
for a given $M_{V_M}$. Fixing $M$ in this way using a generic value $M_{V_M} =1\,{\rm TeV}$, we show the couplings and masses of the gauge boson corresponding to $j=1,2,...,M+1$ for two  sample values of $N$ and $q$ in Fig. \ref{fig2}.
It can be seen that one obtains $M_{V_1} \ge {\cal O}(10^{-3})$ eV for $g_1 \ge 10^{-30}$ almost independent of the values of $N$ and $q$.

Non-observation of supersymmetric particles in the experimental searches so far  typically suggest that the supersymmetry must be broken at scale $\gsim$ few TeV, at least in the visible sector. This can be achieved by introducing the usual soft terms in this framework. The MSSM fields charged under the $U(1)_M$ also receive supersymmetry breaking contribution from the non-vanishing $D_j$. However, such contribution is power suppressed and negligible in comparison to the soft breaking scale. For example, one obtains the largest $D_j$  corresponding to $j=0$ from Eq. (\ref{D_j}), $D_0 \simeq 10^{-5}\, {\rm eV}^2$  for $N = 55$,  $q=0.1$ and $\sqrt{\xi} = 10^{16}$ GeV. Such a small contribution is insignificant for the TeV scale soft masses.

\section{Inflation}
\label{sec:inflation}
Inflation can proceed in a way analogous to the standard $D$-term inflation mechanism proposed in \cite{Stewart:1994ts,Binetruy:1996xj,Halyo:1996pp} (see also \cite{Evans:2017bjs,Domcke:2017xvu,Domcke:2017rzu} for the recent versions). We identify the radial component $\sigma$ of 
\be
X=\frac{1}{\sqrt{2}} \sigma\,e^{i\theta}
\ee
as the inflaton. As discussed earlier, for the inflaton field value $\sigma^2 \ge 2 g_N \xi /\lambda^2_N$ the potential has minimum at $|\phi^\pm_i|=0$ and it is given by 
\be \label{V_tree}
V = \frac{1}{2} \xi^2\,.
\ee
Non-zero $D$-term for the $N^{\rm th}$ $U(1)$, $D_N = \xi$, spontaneously breaks supersymmetry and splits the masses of fermions and bosons residing within $\Phi^\pm_N$. The fermion masses are given by $m_f = \lambda_N |X|$ while the masses of scalars, as can be read off from Eqs. (\ref{VD},\ref{VF}), are given by $m_\pm^2 = \lambda_N^2 |X|^2 \mp g_N \xi$. This splitting in turn generates Coleman-Weinberg correction \cite{Coleman:1973jx} to the tree level potential. The 1-loop correction to the potential can be estimated using
\be \label{CW}
\Delta V= \sum_i \frac{ (-1)^F m_i^4}{64 \pi^2} \ln \left(\frac{m_i^2}{\Lambda^2}\right)\,,\ee
where $i$ runs over the scalars and fermions. $\Lambda$ is the ultraviolet cutoff of the theory which we identify with the reduced Planck scale, $M_P= 1/(8 \pi G)^{1/2}= 2.4 \times 10^{18} \,{\rm GeV}$. The 1-loop corrected effective potential is then given by
\be \label{V_eff}
V_{\rm eff} \equiv V+\Delta V \simeq \frac{1}{2} \xi^2 \left(1+ \frac{g_N^2}{16 \pi^2}\, \ln \left(\frac{\lambda_N^2 \sigma^2}{2 M_P^2}\right) \right)\,,
\ee
for $\lambda^2_N \sigma^2 \gg 2 g_N \xi$. The constant tree level contribution provides the vacuum energy density required to drive inflation and the slow roll is provided by the 1-loop correction.

The values of $g_N$ and $\xi$ can be estimated by fitting the potential in Eq. (\ref{V_eff}) with the observables from inflation models. These observables are minimum number of e-foldings $N$, amplitude of temperature anisotropy $A_s$, spectral index $n_s$ and tensor-scalar ratio $r$. They are given by
\beqa \label{infl_obs}
A_s&=& \frac{V}{24 \pi^2 M_P^4 \epsilon}\,,\label{As} \\
n_s&=&1- 6 \epsilon + 2 \eta\,,\label{ns}\\
N_{\rm CMB}&=&\int H_I\, dt= \int_{\sigma_{c}}^{\sigma_{\rm CMB}} \frac{d\sigma}{M_P \sqrt{2 \epsilon}}\,,\label{NCMB}\\
r&=&16 \, \epsilon\,.\label{r}
\eeqa
Here, $\epsilon$ and $\eta$ are  the slow roll parameters which for the potential in Eq. (\ref{V_eff}) are obtained as
\beqa \label{eps_eta}
\epsilon&=&\frac{M_P^2}{2 V_{\rm eff}^2} \left(\frac{\partial V_{\rm eff}}{\partial \sigma}\right)^2 = 2 \left(\frac{g_N^2}{16 \pi^2}\right)^2 \left(\frac{M_P}{\sigma} \right)^2 \left(1+\frac{g_N^2}{16 \pi^2} \ln\left( \frac{\lambda_N^2 \sigma^2}{2 M_P^2}\right) \right)^{-2} \,,\label{epsilon}\\
\eta&=& \frac{M_P^2}{V_{\rm eff}} \left(\frac{\partial^2 V_{\rm eff}}{\partial \sigma^2}\right) = -\frac{g_N^2}{8 \pi^2} \left(\frac{M_P}{\sigma} \right)^2 \left(1+\frac{g_N^2}{16 \pi^2} \ln\left( \frac{\lambda_N^2 \sigma^2}{2 M_P^2}\right) \right)^{-1}\,. \label{eta}
\eeqa
$N_{\rm CMB}$ is the number of e-foldings between the end of inflation and the time when the CMB modes are exiting the inflationary horizon and $\sigma_{\rm CMB}$ is the value of the inflaton field when the CMB modes are exiting the inflation horizon.  $\sigma_c$ is the critical value of $\sigma$ when inflation ends. Inflation can end in two possible ways. If at some value of $\sigma$, $\epsilon\simeq 1$ then the slow roll phase ends. It is also possible that $\sigma$ reaches the critical value  when the local supersymmetry breaking minimum becomes unstable and the fields roll along the $\phi^+_N$ direction. This critical value is given by  $\sigma_c = \sqrt{2 g_N \xi}/|\lambda_N|$.

\begin{figure}[t!]
\centering
\subfigure{\includegraphics[width=0.45\textwidth]{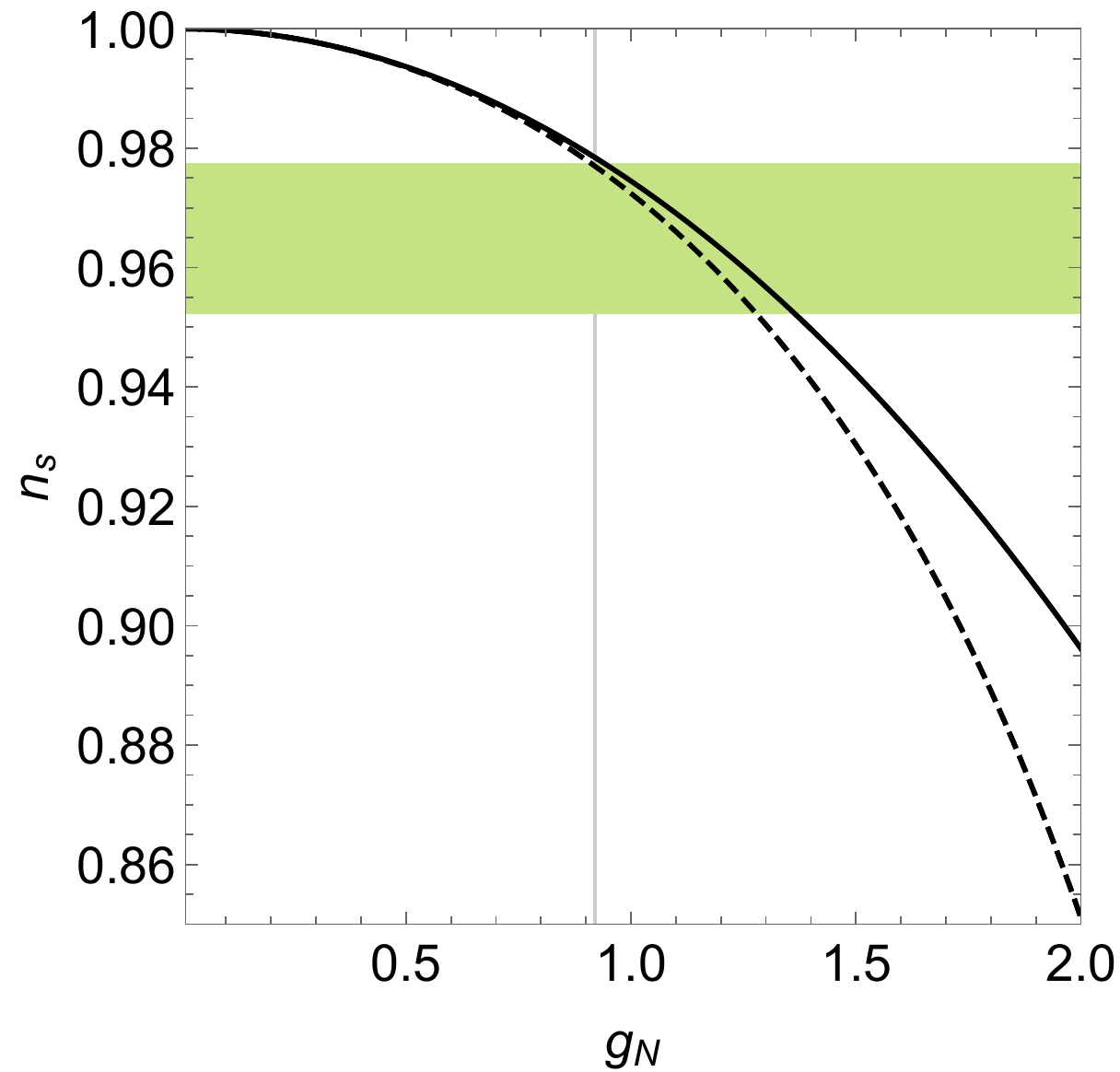}}\hspace{0.7cm}
\subfigure{\includegraphics[width=0.43\textwidth]{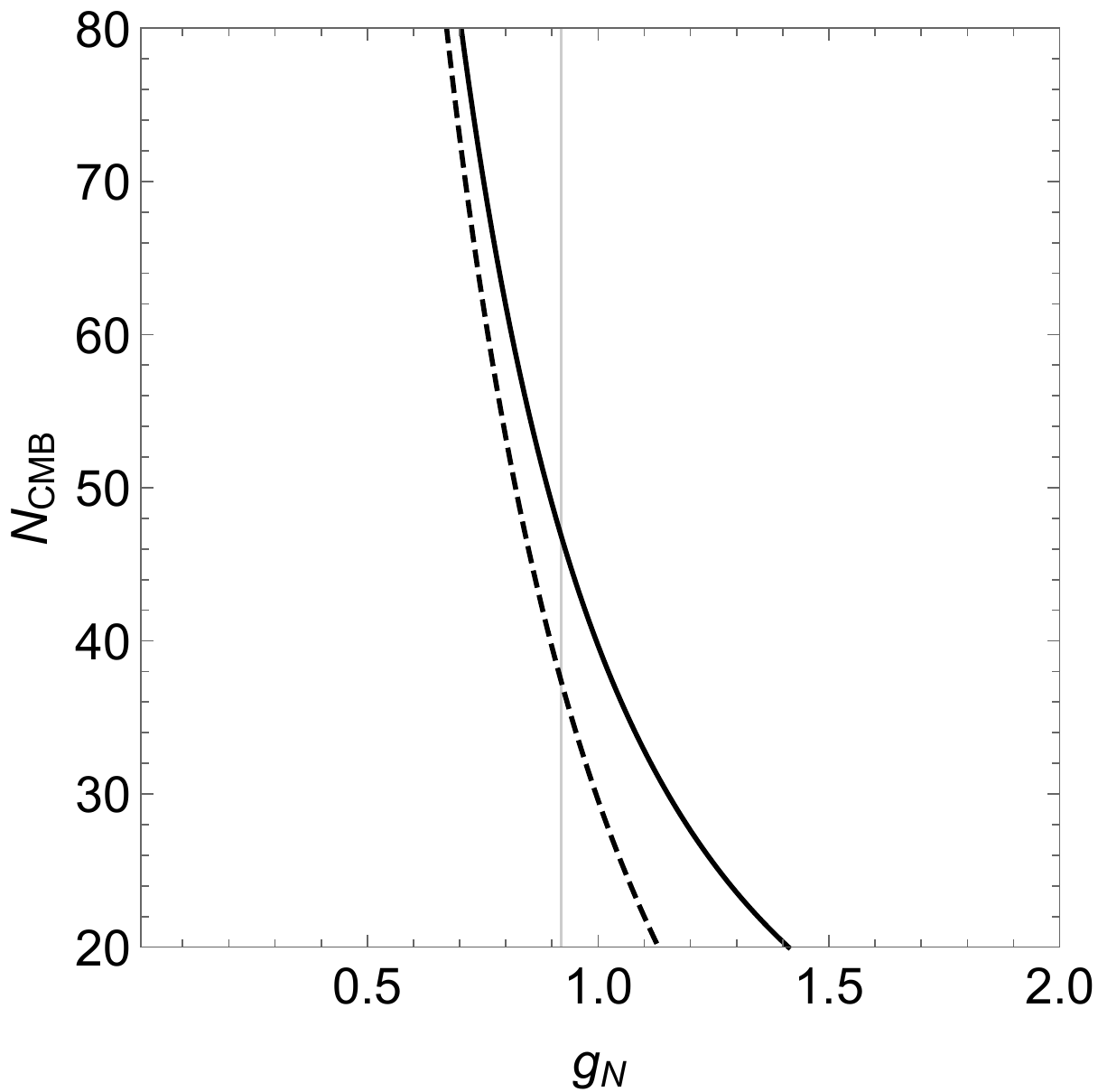}}\\
\subfigure{\includegraphics[width=0.45\textwidth]{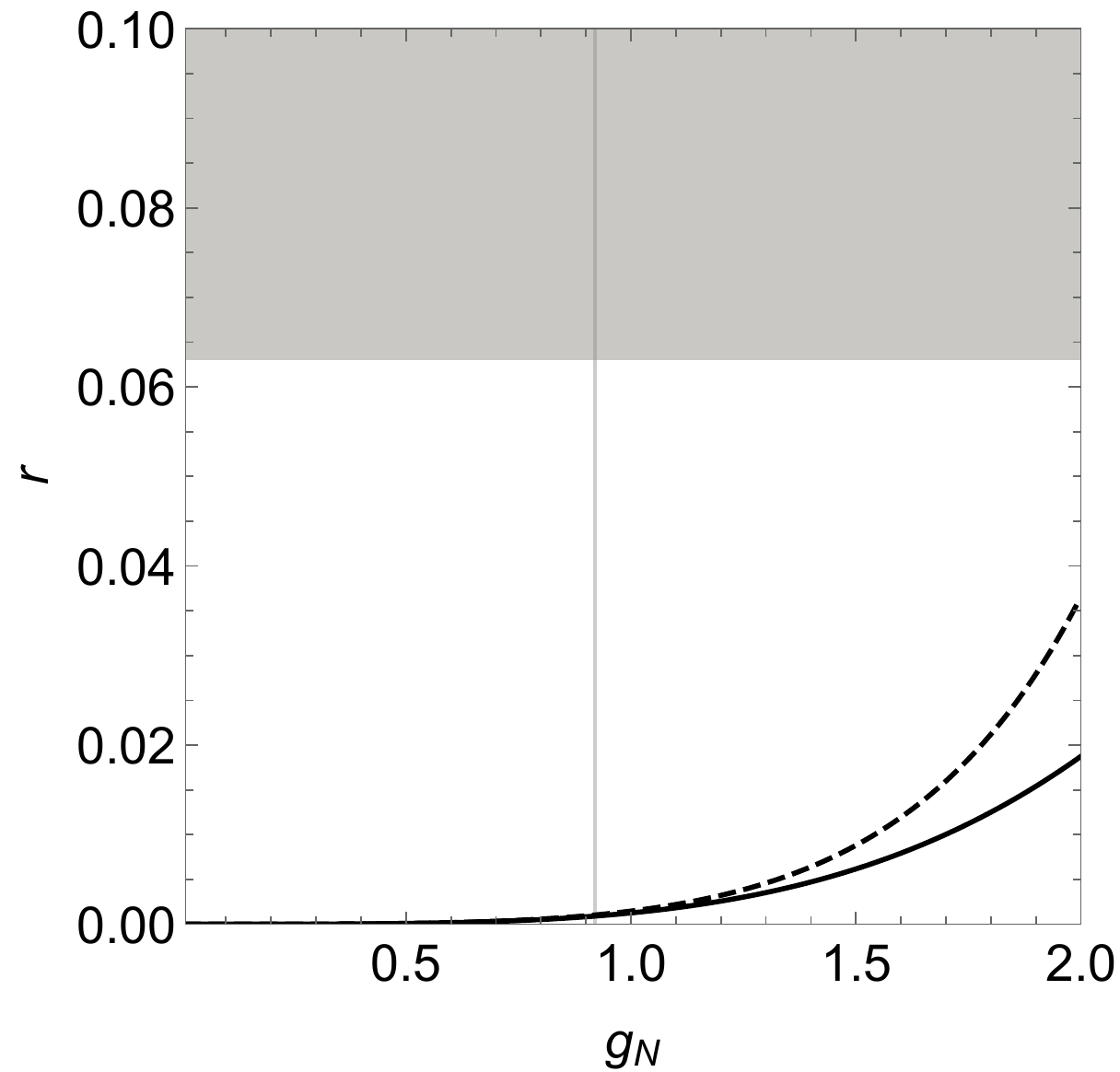}}\hspace{0.2cm}
\subfigure{\includegraphics[width=0.46\textwidth]{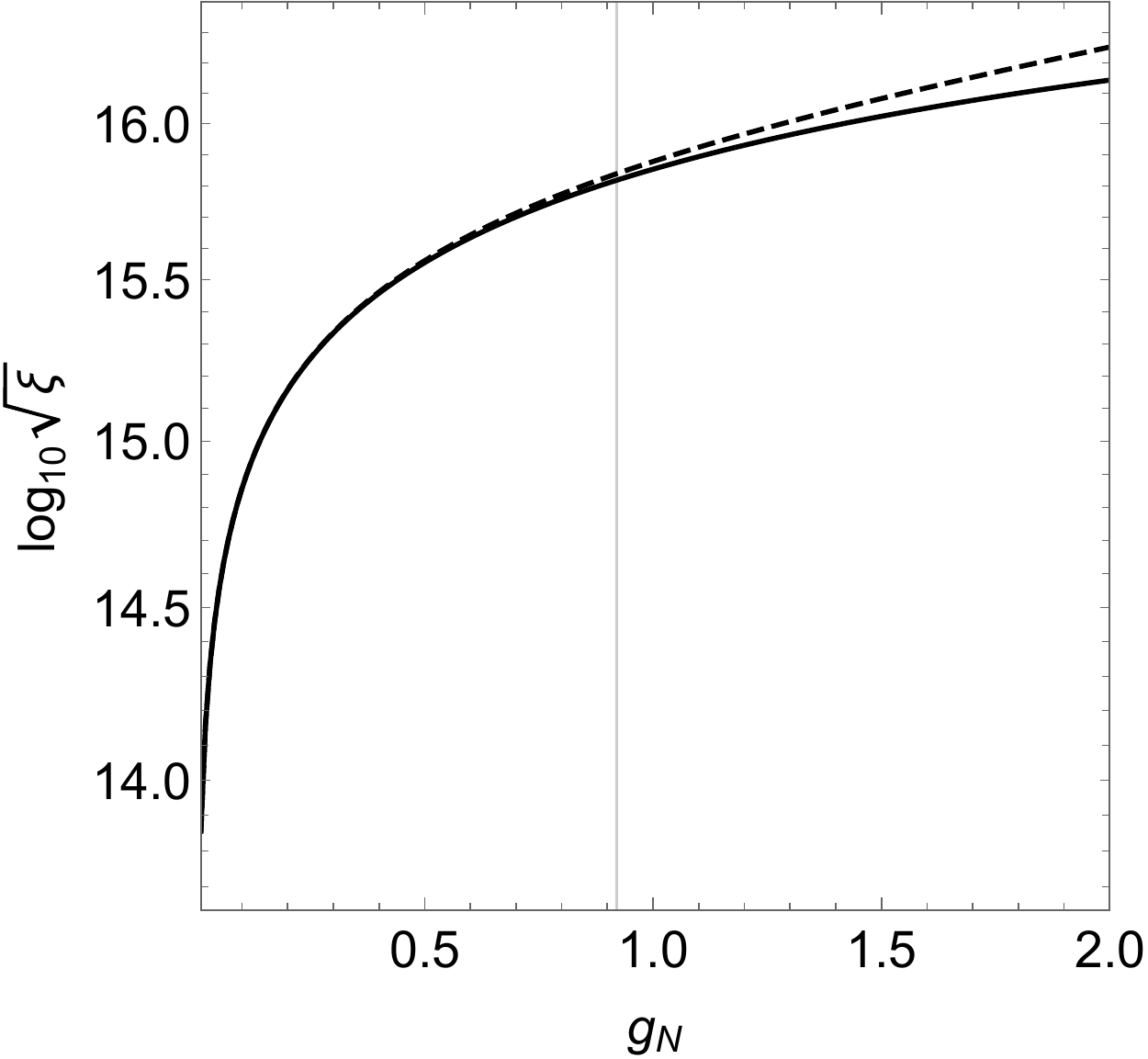}}\\
\caption{The spectral index $n_s$, number of e-foldings $N_{\rm CMB}$, scalar to tensor ratio $r$ and the $\xi$ parameter as functions of $g_N$ as predicted by the model (see the text for details). The dashed (solid) line corresponds to $|\lambda_N| = 10^{-2}$ ($|\lambda_N| = \sqrt{4 \pi}$). The green band in top-left panel indicates $3 \sigma$ range of $n_s= 0.9649 \pm 0.0042$ as measured by Planck 2018 \cite{Akrami:2018odb}. The gray region in the bottom-left panel is excluded by Planck 2018 at 95\% C.L..}
\label{fig3}
\end{figure}

In the following, we assume $\sigma_{\rm CMB} \simeq M_P$ at the start of inflation so that we get ${\cal O}(50)$ e-foldings consistent with the Lyth bound \cite{Lyth:1996im}. The different horizons in the universe can have different values of the inflaton field and the patch of the universe where $\sigma \gtrsim M_P$ will start inflating and will dominate the volume of the universe. The observables given in Eq. (\ref{infl_obs}) are then functions of model parameters $g_N$, $\lambda_N$ and $\xi$. Among these only $A_s$ depends on $\xi$. We scan the values of $g_N$ and $\lambda_N$ to determine all the observables except $A_s$. Parameter $\xi$ is then determined using the value of $A_s$ as measured by Planck 2018, $A_s= (2.099 \pm 0.101) \times 10^{-9}$ \cite{Akrami:2018odb}. The results are displayed in Fig. \ref{fig3}.
We find that $n_s \le 0.9775$, the upper $3 \sigma$ limit as measured by Planck 2018 \cite{Akrami:2018odb}, requires $g_N \ge 0.92$ almost independent of values of $\lambda_N$. For $g_N = 0.92$, $N_{\rm CMB}$ can be as large as 46 for $\lambda_N \simeq \sqrt{4 \pi}$. The greater value of $g_N$ decreases the number of e-foldings considerably. The scalar to tensor ratio remains  well below the upper bound, $r < 0.062$. For $g_N = 0.92$, the value of $\xi$ fixed from $A_s$ is determined as $\sqrt{\xi} \simeq 6.3 \times 10^{15}$ GeV.

The inflation parameters $A_s$ and $n_s$ measured by Planck 2018 \cite{Akrami:2018odb} are reported for the scale $k=0.05 \,{\rm Mpc^{-1}}$. The mode with co-moving wavenumber $k$ exits the inflation horizon when the physical length scale of the perturbation $a_k/k$ is the size of the horizon $H_I^{-1}$, i.e when $a_k  =k  H_{I}^{-1}$. Therefore the number of e-foldings $N(k)$ before the end of inflation when a given mode $k$ leaves the inflation horizon is given by \cite{Liddle:1993fq,Liddle:2003as,Dodelson:2003vq}
\beqa
e^{N(k)}= \frac{a_{end}}{a_k}= a_{end} \frac{H_I}{k}= \left(\frac{H_I}{k}\right)\left(\frac{a_{end}}{a_{reh}}\right)\left(\frac{a_{reh}}{a_{eq}}\right)\left(\frac{a_{eq}}{a_{0}}\right)\,,
\eeqa
which implies
\be \label{Nk}
N(k)= \ln \left(\frac{H_I}{k}\right) + \frac{1}{3}  \ln \left(\frac{\rho_{reh}}{\rho_{end}}\right)+\frac{1}{4} \ln \left(\frac{\rho_{eq}}{\rho_{reh}}\right)+ \ln \left(\frac{a_{eq}}{a_0}\right)\,.
\ee
Here, we have assumed that the at the end of inflation the inflaton oscillates at the bottom of the potential and the energy density falls as $\rho\sim a^{-3}$ and then the universe reheats due to the coupling of inflaton with the SM fields. In Eq. (\ref{Nk}),  $H_I= (V/3 M_P^2)^{1/2}$ is the Hubble parameter during inflation where the potential $V\simeq \xi^2/2$.  With $\xi= (6.3 \times 10^{15}\, {\rm GeV})^2$  we obtain the value of $H_I= 6.65 \times 10^{12}\, {\rm GeV}$. The ratio $a_0/a_{eq}= 3450$ and $T_{eq}= 0.81$ eV.  Using these parameters in Eq. (\ref{Nk}) and taking $N(k=0.05 \, {\rm Mpc^{-1}} )\simeq 46 $  the reheat temperature turns out to be 
$T_{reh}=10^6 \,{\rm GeV}$. Reheating at the end of inflation takes place when the $X$ particles decay into $\phi^{\pm}_N$ and gauge bosons.

\section{Constraints from long-range forces}
\label{sec:constraints}
As discussed in the previous section, a viable inflation within this framework requires
\be \label{g_N-xi}
g_N \simeq 1\,~~{\rm and}~~\sqrt{\xi} \simeq 10^{16}\, {\rm GeV}\,.
\ee
Substitution of the above in Eqs. (\ref{MZp},\ref{g'},\ref{M}) determines the allowed values of $g^\prime$ and $M_{Z^\prime}$ as function of CW parameters $N$, $k$ and $q$. $M_{Z^\prime}$ also depends on $v_\chi \lesssim v_k$. To be more specific, we associate $v_\chi$ with the breaking scale of $U(1)_k$ by assuming
\be \label{vchi_vk}
v_{\chi} \simeq v_k = q^{\frac{N-k}{2}}\,\sqrt{\frac{\xi}{g_N}}\,, \ee
where the second equality results from Eq. (\ref{minimum}). This, along with the above values of $g_N$ and $\xi$, leads to
\be \label{k}
k = 2 \left(\frac{\log_{10} \left(M_{Z^\prime}/{\rm eV}\right) -25}{\log_{10} q} \right) -N\,,
\ee
from Eq. (\ref{MZp}). Similarly, substitution of the values of $g_N$, $\xi$ and $M_{V_M} = 1$ TeV in Eqs. (\ref{g'},\ref{M}) implies
\be \label{N}
N = \frac{\log_{10} g^\prime - 26}{\log_{10} q}\,.
\ee
Desired value of $g^\prime$ and $M_{Z^\prime}$ can therefore be obtained by choosing appropriate values of $N$ and $k$ for a given $q$. The ratio $k/N$ however does not depend on the value of $q$ and it can be constrained once the nature of the SM current $J^\mu$ is fixed. We do this by identifying $U(1)_M$ with $L_\mu-L_\tau$, $L_e - L_\tau$ and $B-L$ symmetries.

\subsection{$L_\mu - L_\tau$}
\begin{figure}[ht!]
\centering
\subfigure{\includegraphics[width=0.90\textwidth]{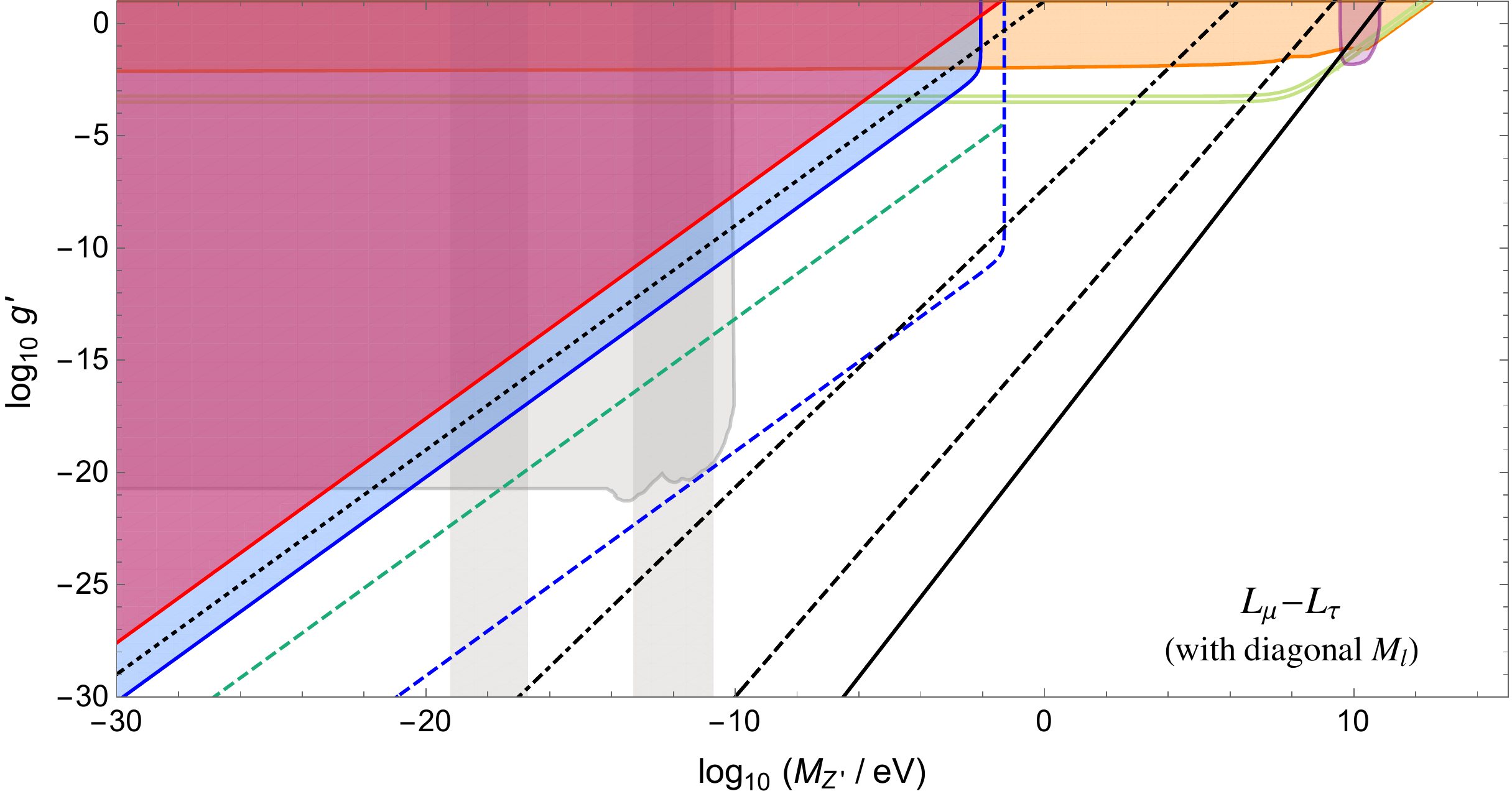}}\\ 
\subfigure{\includegraphics[width=0.90\textwidth]{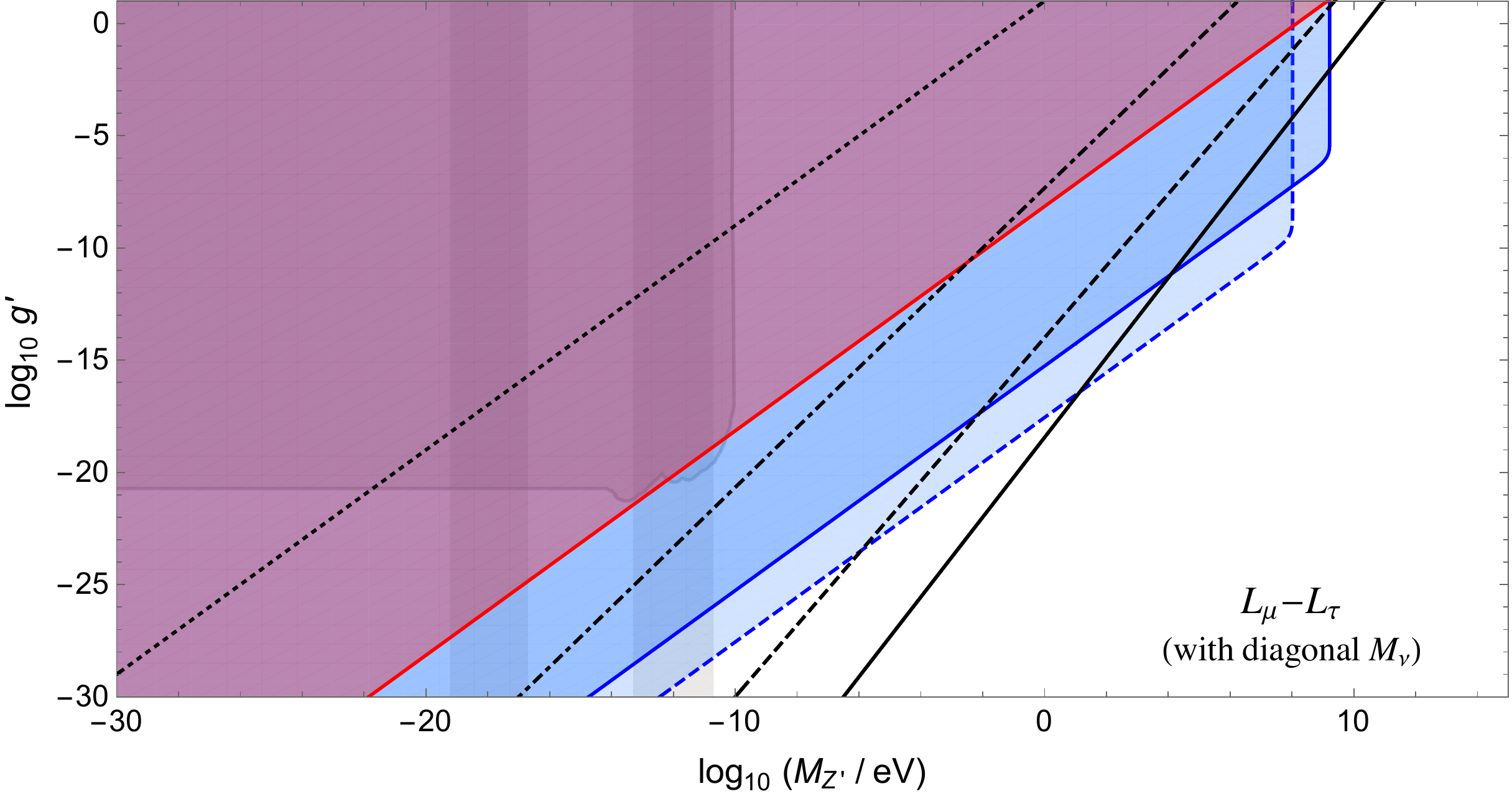}}
\caption{Constraints on $g^\prime$ and $M_{Z^\prime}$ for $U(1)_M = L_\mu-L_\tau$ with diagonal $M_l$ ($M_\nu$) in the top (bottom) panel. The black lines represent model predicted correlations for $k/N=1/8$ (solid), $1/4$ (dashed), $1/2$ (dot-dahsed) and $k/N=1$ (dotted). The red shaded region is excluded by unitarity limit \cite{Dror:2020fbh}. The vertical gray bands indicate the range of $M_{Z^\prime}$ disfavoured by black hole superradiance \cite{Baryakhtar:2017ngi} while the gray region enclosed by grey line is excluded by neutron star binaries \cite{Poddar:2019wvu,Dror:2019uea}. Shaded in orange (purple) is the region disfavoured by neutrino trident production \cite{Altmannshofer:2014pba} (LHC \cite{Aad:2014wra,CMS:2012bw}) constraints. The region enclosed between two green contours is favoured by muon $g-2$ at $2 \sigma$ \cite{Baek:2001kca,Aoyama:2020ynm}. Limit set by BBN due to neutrino annihilation is shown by dashed green line \cite{Dror:2020fbh}. The regions enclosed by solid and dashed blue lines in the top panel are disfavoured by laboratory and cosmological constraints on neutrino decays \cite{Dror:2020fbh}, respectively. The same in the lower panel are excluded by upper limits on ${\rm BR}[\tau \to \mu Z^\prime]$ \cite{Albrecht:1995ht}  and  ${\rm BR}[\mu \to e Z^\prime]$ \cite{Bayes:2014lxz}, respectively. All the constraints are at $95 \%$ C.L..}
\label{fig4}
\end{figure}

We discuss here various constraints which are used to restrict the parameter space in case of $U(1)_{L_\mu-L_\tau}$ symmetry and its consequences for the CW setup considered here. The second and the third generation of leptons are charged under $U(1)_M$ with charges $+1$ and $-1$, respectively but the first generation is neutral. Thus objects containing electrons do not experience the $L_\mu-L_\tau$ forces and the conventional method used to constrain fifth force do not apply\footnote{These constraints become meaningful if $Z^\prime$ has a mass mixing \cite{Davoudiasl:2011sz,Heeck:2010pg} with the ordinary $B$ boson. We assume that such mixing is not present.}. But the muon rich astrophysical sources like neutron star binaries can provide  significant constrain on ultra-light gauge boson. Emission of an ultra-light gauge boson of $L_\mu-L_\tau$ causes a fast decay in the orbital period of a pulsar binary. The observed orbital periods have  been used in \cite{Poddar:2019wvu,Dror:2019uea} to constrain the mass and couplings of the $L_\mu-L_\tau$ gauge boson. The masses $M_{Z^\prime}$ below around $10^{-10}$ eV are constrained in this way. This constraint is shown in Fig. \ref{fig4} as a grey region enclosed by grey line \cite{Dror:2019uea}. 

$M_{Z^\prime}$ above $10^{-10}$ eV but below $m_{\nu_i}$ are constrained  from the  invisible neutrino decays $\nu_i \to \nu_j + Z^\prime$. The strongest limit on the neutrino lifetime comes from the structure formation in the early universe \cite{Hannestad:2004qu,Hannestad:2005ex} through CMB observation by Planck 2018 \cite{Escudero:2019gfk}. This constraint is shown in the upper panel of Fig. \ref{fig4}. This limit is however cosmological model dependent and it does not apply if neutrinos disappear before recombination epoch through additional decay channels, see for example \cite{Beacom:2004yd}.  A less stringent but more robust bound on invisible neutrino decays come from the laboratory data on non-observation of the $\nu_2$ decays \cite{Aharmim:2018fme} and it disfavours the region shaded by blue in the top panel of Fig. \ref{fig4}. Higher mass range can be constrained from various other considerations. The coupling of the longitudinal $Z^\prime$ to neutrinos goes as $g^\prime m_\nu / M_{Z^\prime}$ and could lead to unitarity violation. This provides a limit \cite{Dror:2020fbh} on $g^\prime$ for a larger mass range in $M_{Z^\prime}$ as displayed in Fig. \ref{fig4}. Neutrino trident production \cite{Altmannshofer:2014pba} also provides a  strong complementary limits on $g^\prime$ for $M_{Z^\prime} > m_\nu$. This is shown as a region shaded in orange in Fig. \ref{fig4}. As it can be seen, the various constraints still allow $M_{Z^\prime} \sim 10^{-14}$-$10^{-18}$ eV and $g^\prime \sim 10^{-26}$-$10^{-30}$. This region can be obtained in the CW framework for $k=N/2$ curve displayed in Fig. \ref{fig4}. For example, $N=52$, $k=26$, $q=0.1$ give $g^\prime \sim 10^{-26}$ and $M_{Z^\prime}\sim 10^{-14}$ eV.

The constraints from the invisible neutrino decay and neutrino trident productions used above  implicitly assume that the charged lepton mass matrix preserves $L_\mu-L_\tau$ symmetry 
and is diagonal. One can consider alternative case with a diagonal neutrino mass matrix. Leptonic mixing then requires a non-diagonal charged lepton mass matrix.  In this case, the invisible neutrino decay will be absent at the tree level and constraints from the neutrino trident production would also change. But the charged leptons in this case  have flavour violating couplings and $\mu$ and $\tau$ could decay into ultra-light $Z^\prime$ in this case. We estimate these decays following a similar formalism used in \cite{Dror:2020fbh} for neutrinos. The $l_i \to l_j + Z^\prime$ decay width can be obtained as 
\be \label{l_decay}
\Gamma [l_i \to l_j\, Z^\prime ] = \frac{1}{32 \pi m_{l_i}^3}\, \frac{{g^\prime}^2}{M_{Z^\prime}^2}\,  |Q_{ij}|^2 \left( m_{l_i}^2 - m_{l_j}^2\right)^2\, \lambda^{1/2}\left(m_{l_i}^2, m_{l_j}^2, M_{Z^\prime}^2\right)\,, \ee
where, $\lambda(x,y,z) = x^2 + y^2 + z^2 - 2xy -2 yz- 2zx$ and $Q = U_{\rm PMNS}\, \hat{Q}\, U_{\rm PMNS}^\dagger$ with $\hat{Q} = {\rm Diag.}(0,1,-1)$ for the underlying case.  We use the results of latest fit \cite{Esteban:2020cvm} of neutrino oscillation data to determine the $U_{\rm PMNS}$ matrix and estimate the branching ratios for the decays, $\tau \to \mu + Z^\prime$ and $\mu \to e + Z^\prime$. The upper limits ${\rm BR}[\tau \to \mu\, Z^\prime] < 5 \times 10^{-3}$ \cite{Albrecht:1995ht} and ${\rm BR}[\mu \to e\, Z^\prime] < 5.8 \times 10^{-5}$ \cite{Bayes:2014lxz} are then used to constrain $g^\prime$ and $M_{Z^\prime}$ in the lower panel of Fig. \ref{fig4}. As can be seen, these constraints are more powerful and exclude $k/N \gsim 1/4$ leaving no room for the long-range interactions.

$L_\mu-L_\tau$ symmetry has also been evoked to explain the current discrepancy between theoretically predicted muon anomalous magnetic moment $(g-2)_\mu$ and its experimental value. This can be resolved if $g^\prime \simeq 10^{-3}$ and $M_{Z^\prime} \simeq {\cal O}$(MeV) as seen  from the top panel of Fig. \ref{fig4}. Such values can be accommodated in the proposed CW by taking $N=29$, $k=9$ for $q=0.1$ as can be seen from Eqs. (\ref{k},\ref{N}). For this choice, one obtains $M=3$ for which $M_{V_M}\sim$ TeV as explained earlier.

\subsection{$L_e - L_\mu$}
\begin{figure}[ht!]
\centering
\subfigure{\includegraphics[width=0.90\textwidth]{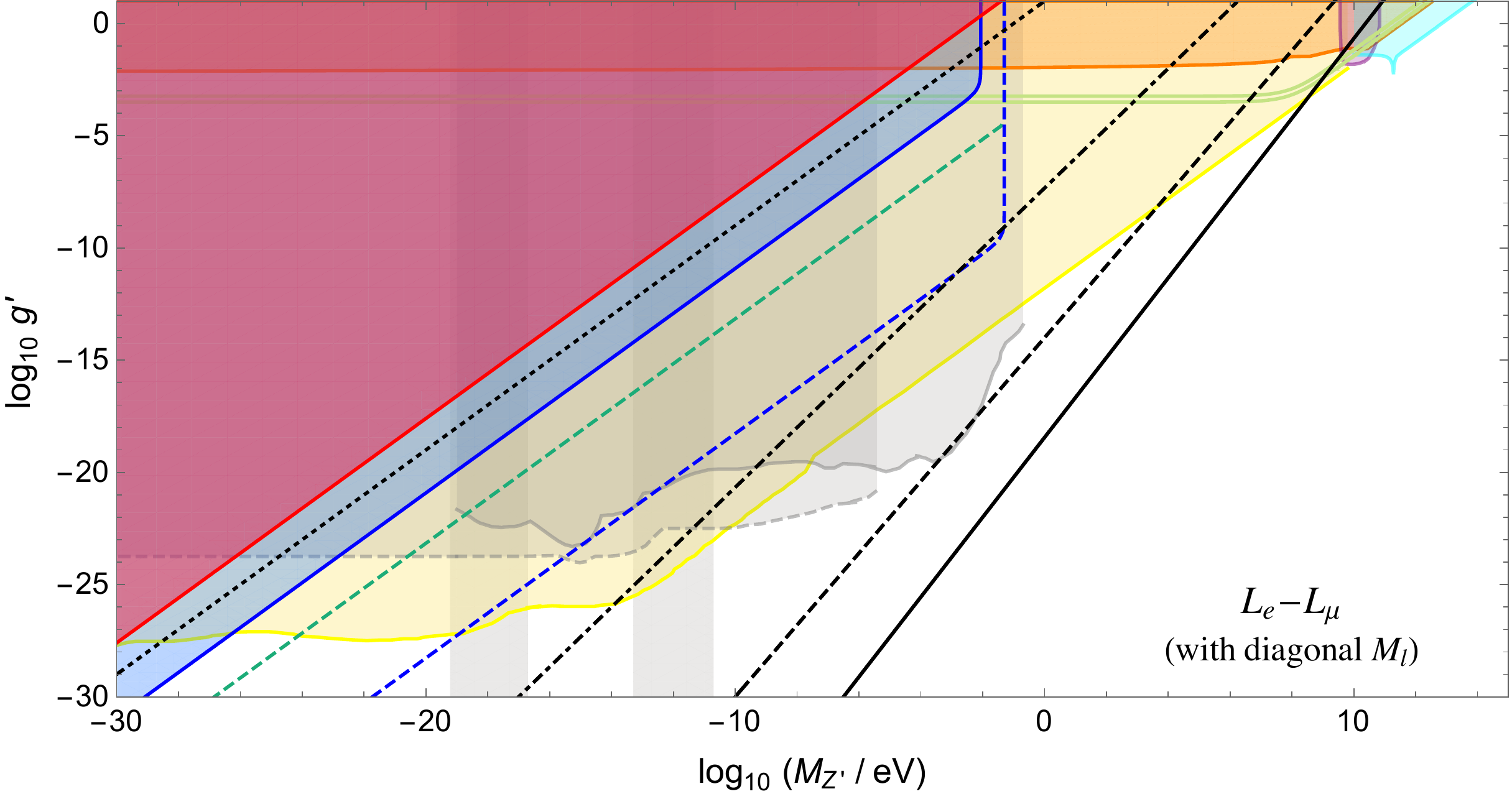}}\\ 
\subfigure{\includegraphics[width=0.90\textwidth]{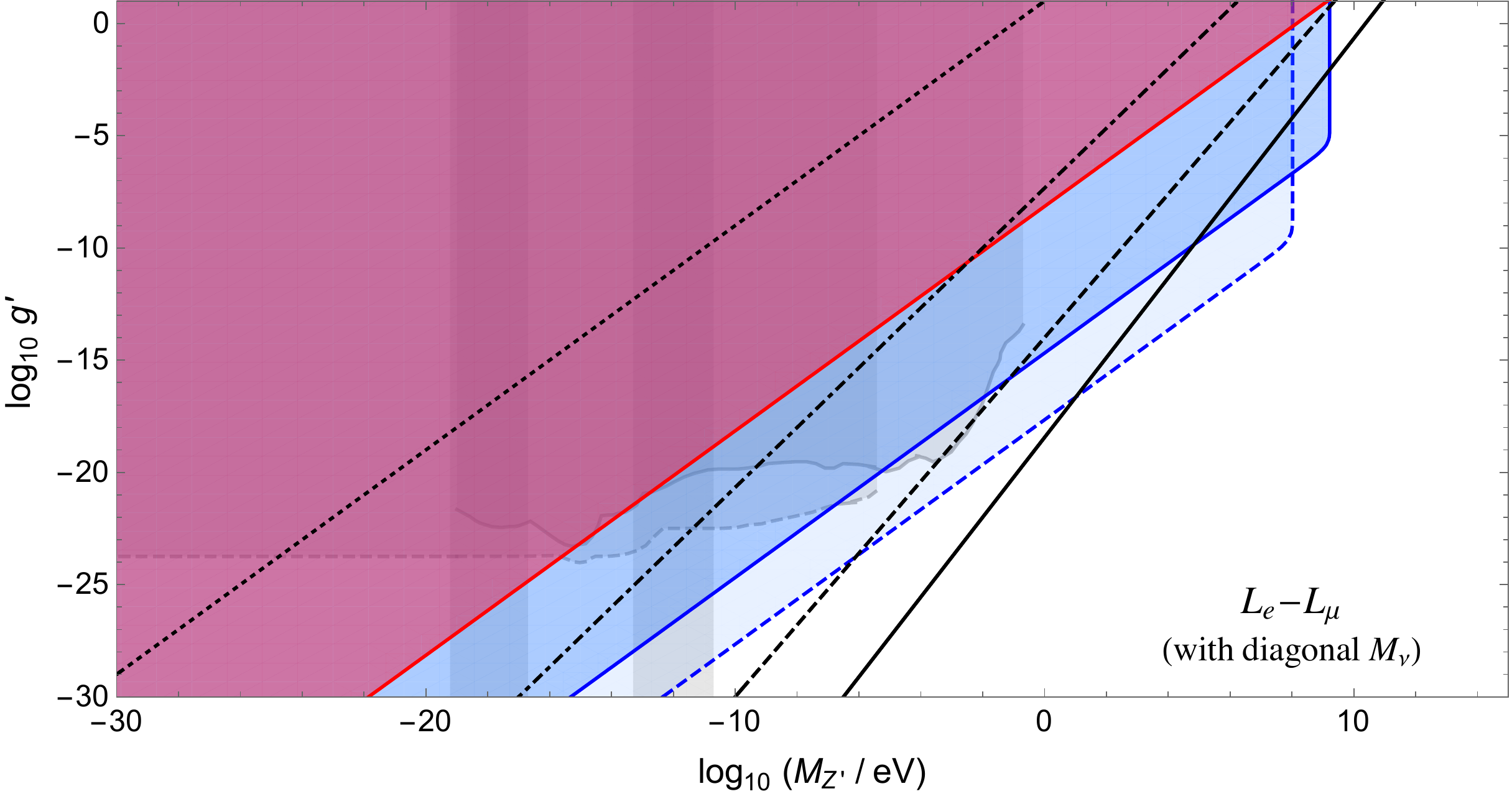}}
\caption{Constraints on $g^\prime$ and $M_{Z^\prime}$ for $U(1)_M = L_e-L_\mu$ with diagonal $M_l$ ($M_\nu$) in the top (bottom) panel. The region shaded in yellow is disfavoured by matter effects in neutrino oscillation \cite{Wise:2018rnb,Bustamante:2018mzu}. Grey regions enclosed by solid (dashed) grey contours are excluded by experimental test of violation of equivalence principle (fifth force) \cite{Wise:2018rnb}. The region shaded in cyan is excluded by LEP \cite{Schael:2013ita}. All the other details are same as discussed in the caption of Fig. \ref{fig4}.}
\label{fig5}
\end{figure}

In this case, the $Z^\prime$ boson has gauge interactions with the first generation leptons and therefore the stringent constraints on $g^\prime$-$M_{Z^\prime}$ arise from the matter effects in neutrino oscillations \cite{Wise:2018rnb} and from the precision tests of gravity. These are more powerful in comparison to the limits from neutrino decays as can be seen from the top panel in Fig. \ref{fig5}. The constraints still allows some room for the long-range $L_e-L_\mu$ interaction corresponding to $g^\prime \simeq 10^{-27}$ and $M_{Z^\prime} \simeq 10^{-15}$ eV which can be incorporated in the proposed CW set-up with $k = 27$ and $N=53$ for $q=0.1$.  

We also consider the constraints from the charged lepton decays which arise if $L_e-L_\mu$ is broken by the charged lepton mass matrix and the neutrino mass matrix is diagonal. The same procedure as outlined in the previous subsection is followed but with $\hat{Q} = {\rm Diag.}(1,-1,0)$ in Eq. (\ref{l_decay}). The flavour violating charged lepton decays provide the most stringent constraints and entirely exclude $k/N \ge 1/4$ as displayed in the bottom panel in Fig. \ref{fig5}.  Most of the constraints and results discussed in this subsection are also applicable to the $L_e - L_\tau$ type of interactions.

\subsection{$B-L$}
\begin{figure}[ht!]
\centering
\subfigure{\includegraphics[width=0.90\textwidth]{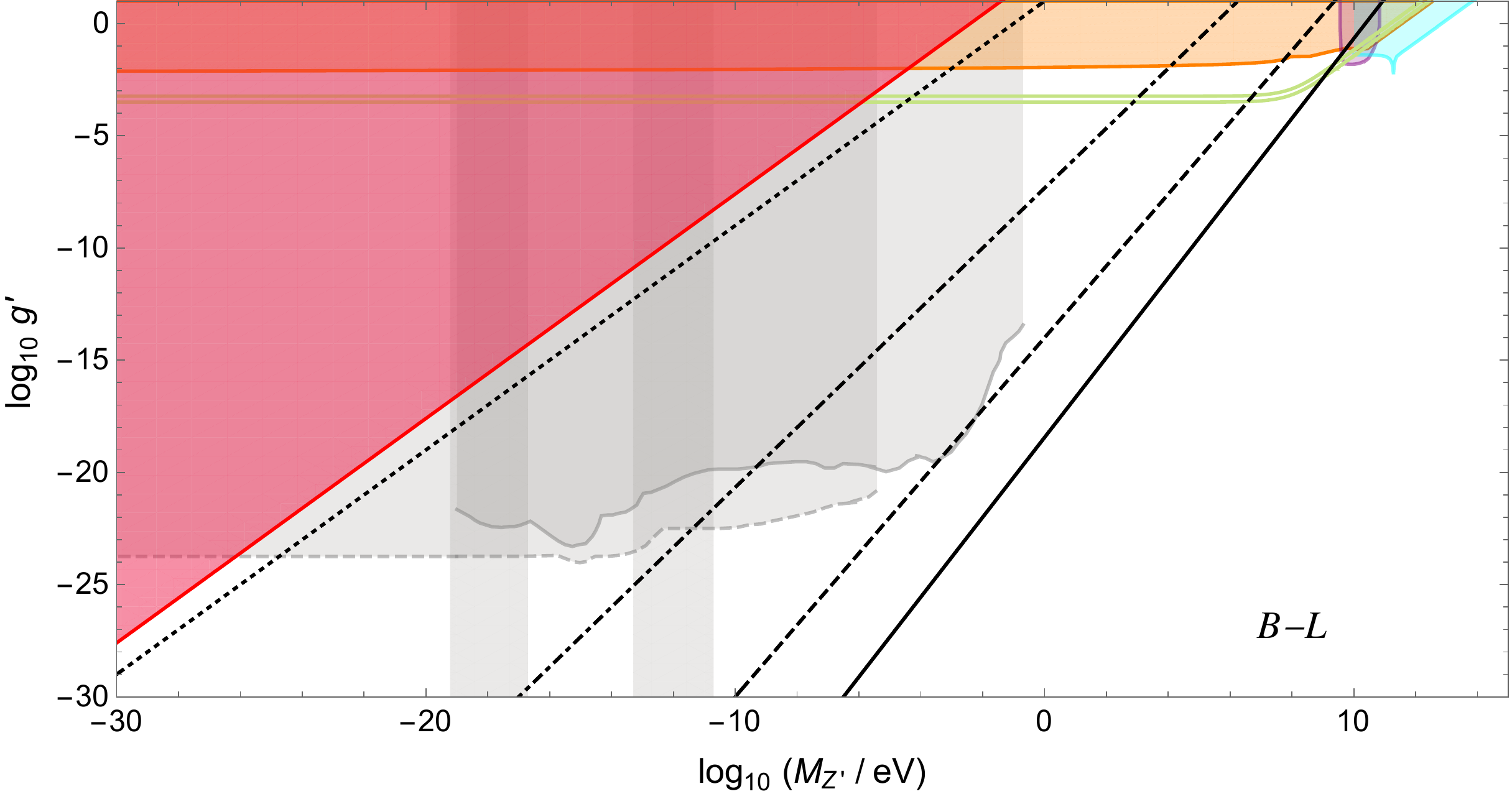}}
\caption{Constraints on $g^\prime$ and $M_{Z^\prime}$ for $U(1)_M = B-L$. Grey regions enclosed by solid (dashed) grey contours are excluded by experimental test of violation of equivalence principle (fifth force) \cite{Wise:2018rnb}. The region shaded in cyan is excluded by LEP \cite{Schael:2013ita}. All the other details are same as discussed in the caption of Fig. \ref{fig4}. We do not show various other constraints applicable in the range of $M_{Z^\prime}$ in 1 eV - 10 GeV, see \cite{Heeck:2014zfa,Ilten:2018crw} for their details.}
\label{fig6}
\end{figure}

For $U(1)_M = B-L$, the gauge interactions involving $Z^\prime$ are flavour universal. The dominant constraint on the ultra-light $Z^\prime$ comes from the experiments testing the validity of equivalence principle and existence of the fifth force as discussed before. These constrains allows $g^\prime \le 10^{-24}$ and $M_{Z^\prime} \simeq 10^{-14}$-$10^{-16}$ eV which can be obtained within the proposed CW set-up for $N \ge 55$ and $k/N \sim 1/2$ as can be seen from Fig. \ref{fig6}.  The unitarity constraint shown in red color in Fig. \ref{fig6} is  applicable if neutrinos break the lepton number. Also, we do not show various other constraints applicable in the range of $M_{Z^\prime}$ in 1 eV - 10 GeV (see \cite{Heeck:2014zfa,Ilten:2018crw} for example) which is not of importance in the present context.

\section{Summary}
\label{sec:concl}
Extension of the minimal supersymmetric SM by a $U(1)$ gauge group is known to lead to a successful description of inflation through a large FI term $\xi$. We have incorporated this mechanism of the $D$-term inflation into a broader framework containing $N+1$ different $U(1)$ gauge groups coupled with each other through Higgs fields in a clockwork fashion. As discussed here, this offers an exciting possibility of unifying the large scale inflation and long-range interactions mediated by an ultra-light gauge boson, and in-turn explains forty orders of magnitude difference between the scales without relying on any unnatural parameter. Such long-range forces can be minimally incorporated in the SM by gauging $U(1)$ symmetry corresponding to the difference $L_i-L_j$ of individual lepton number. Such $U(1)$ symmetries form a part of the underlying CW framework. Both tiny mass $\sim 10^{-15}$ eV of the gauge boson and its extremely weak coupling $g^\prime \sim 10^{-26}$ to the SM fields arise from the underlying CW mechanism. Simultaneously, it also allows ${\cal O}$(1) coupling $g_N$ needed for $D$-term inflation. As discussed at length in section \ref{sec:inflation}, various conditions required for successful inflation can be met within the present scenario. 

We have considered in detail three specific cases of the long-range forces generated by the $L_\mu-L_\tau$, $L_e-L_\mu$ and very light $B-L$ gauge bosons. We have collected most of the relevant astrophysical, cosmological and terrestrial constraints in these scenarios and shown that the values of parameters $g^\prime$ and $M_{Z^\prime}$ surviving after these constraints can be understood within this framework. 

The proposed  CW mechanism is not restricted to the description of the long-range forces. The same setup also allows a heavier $M_{Z^\prime}$ (see Fig \ref{fig4}) or heavier excitations of the lowest mass state (see Fig. \ref{fig2}) with stronger couplings to SM fields than the ones required for the long-range forces. They can be used as an explanation of some other physical situations along with inflation. We considered here a specific case of $M_{Z^\prime} \sim {\cal O}$(MeV)  with  coupling $g^\prime \sim 10^{-3}$  that arise in the CW and provide a possible explanation of the long-standing muon $(g-2)$ discrepancy.

\section*{Acknowledgements}
We are grateful to Pierre Fayet for pointing out an error in the previous version of the manuscript.  The work of KMP is partially supported by research grant under INSPIRE Faculty Award (DST/INSPIRE/04/2015/000508) from the Department of Science and Technology, Government of India.

\bibliography{references}
\end{document}